\documentclass{aa}  
\usepackage{natbib}

\usepackage{hyperref} 
\expandafter\let\csname ver@showkeys.sty\endcsname\relax

\usepackage{bm}
\usepackage{graphicx}
\usepackage{txfonts}
\usepackage{float}
\usepackage{stfloats}
\usepackage[section]{placeins}
\usepackage{cuted}
\usepackage{caption}

\begin{document}

   \title{Asteroseismology and Buoyancy Glitch Inversion with Fourier Spectra of Gravity Mode Period Spacings}

   \author{Zhao Guo
     }
   \institute{Institute of Astronomy (IvS), Department of Physics and Astronomy, KU Leuven, Celestijnenlaan 200D, 3001 Leuven, Belgium\\
              \email{zhao.guo@kuleuven.be}
             }

   \date{Received August 15, 2025; accepted Oct 16, 2025}

 
  \abstract
   { }
   {We investigate the small, quasi-periodic modulations seen in the gravity-mode period spacings ($\Delta P_k$) of pulsating stars. These `wiggles' are produced by buoyancy glitches - sharp features in the buoyancy frequency ($N$) caused by composition transitions and the convective–radiative interface. 
   
   }
   {Our method takes the Fourier transform of the period-spacing series, $FT(\Delta P_k)$ as a function of radial order $k$. We show that $FT(\Delta P_k)$ traces the radial derivative of the normalized glitch profile $\delta N/N$ with respect to the normalized buoyancy radius; peaks in $FT(\Delta P_k)$ therefore pinpoint jump/drop locations in $N$ and measure their sharpness. We also note that the Fourier transform of relative period perturbations (deviations from asymptotic values), $FT(\delta P/P)$, directly recovers the absolute value of glitch profile $|\delta N/N|$ , enabling a straightforward inversion for the internal structure.}
   { The dominant $FT(\Delta P_k)$ frequency correlates tightly with central hydrogen abundance ($X_c$) and thus with stellar age for Slowly Pulsating B-stars, with only weak mass dependence. Applying the technique to MESA stellar models and to observed slowly pulsating B-stars and $\gamma$ Dor pulsators, we find typical glitch amplitudes $\delta N/N \lesssim 0.01$ and derivative magnitudes $ \lesssim 0.1$, concentrated at chemical gradients and the convective boundary. This approach enables fast, ensemble asteroseismology of g-mode pulsators, constrains internal mixing and ages, and can be extended to other classes of pulsators, with potential links to tidal interactions in binaries.
      } 
   {}

   \keywords{Stars: evolution --
                Stars: interiors --
                Asteroseismology --
                Methods: analytical
               }

   \maketitle
%

\section{Introduction}

Stars, being self-gravitating fluids (plasmas), mostly contain stratified
regions that are capable of supporting internal gravity 
waves \citep{Sut10}. In those regions, buoyancy acts as the restoring force \citep{Tur73}, 
enabling gravity waves to propagate and form normal modes (g modes), thereby 
allowing us to probe the deep stellar interior. 
In the asymptotic limit, high-order g-modes form 
nearly uniformly spaced sequences in pulsation periods \citep{Tas80}, 
with a characteristic asymptotic period spacing 
that depends sensitively on the integral of the Brunt-Väisälä
 frequency (buoyancy frequency) through the propagation cavity.

Local sharp features such as the base of the 
convective envelope, He II ionization zone, and the convetive-core
boundary, can act as glitches to acoustic modes and induce variations
 in the small frequency separations \citep{Rox94,Bas97}.
Similarly, buoyancy glitches such as composition-transition 
regions can lead to deviations from the 
regular period-spacing pattern. 
Investigating these glitche-induced seismic signatures enables one
to gain direct access to sharp structural features 
and mixing processes in the stellar interior \citep{Cun24}.

Observationally, gravity-mode period-spacing patterns have now been 
detected across a wide range of pulsating stars. 
For $\gamma$ Doradus stars, photometry from {\it Kepler} revealed clear g-mode sequences, such as those in KIC 11145123 \citep{Kur14} 
 and 611 stars in \citet{Li19}.
Slowly Pulsating B (SPB) stars extend this diagnostic to higher masses:
\citet{Deg10} discovered period spacing deviates from a uniform spacing;
\citet{Pap14, pap15} reported period spacings in SPBs
KIC 10526294 and KIC 7760680, followed 
by detailed seismic modeling by \citet{Mor15, Mor16}.
\citet{Ped18, Ped21} performed detailed seismic modeling 
to infer mixing profiles for 26 SPB systems. 
Similar modeling work includes \citet{Wu18, Wu20} and \citet{Sze18}. 
\citet{Shi23} identified 286 new SPB candidates using
TESS, LAMOST, and Gaia surveys. 

Compact pulsators such as white dwarfs \citep{Alt10, Cor19} and subdwarf B stars (sdB), also exhibit g modes 
\citep{Bra92,Cha00,Uzu17,Guy25}.
Recent analyses of buoyancy glitches in sdB stars are presented in \citet{Cun24}. 
Red-giant stars \citep{Mos14,Cun19,Ong21} also exhibit gravity modes of mixed 
character.

Theoretical studies of g-mode pulsation periods have focused on incorporating rotational effects via the traditional approximation of 
rotation \citep{Tow13,Bou13} and accounting for
 centrifugal deformation \citep{Hen21} and magnetic fields \citep{Rui24, Lig24}.

Oscillatory signatures in the period-spacing patterns have been 
studied via the Fourier transform \citep{Wu18} and they found
a tight relation between the frequency of variation $f_{\Delta P}$
and central hydrogen fraction $X_c$. A diagram ($f_{\Delta P}$ vs. $\Pi_0$) 
analogous to the classical C-D diagram ($\Delta\nu$ vs $\delta\nu_{02}$) for acoustic modes in
solar-like oscillators was constructed. \citet{Zha23} 
and \citet{Hat23} studied the prescribed buoyancy-glitch
expressions and the induced period spacing variations.

Inspired by \citet{Wu18}, we extend the work of \citet{Mig08}
 and \citet{Zha23} and demonstrate that the Fourier spectrum of the gravity-mode period spacings
 $FT(\Delta P_k)$, can be used to probe the Brunt--V\"ais\"al\"a (Brunt for short) profile jump/drop points
and to reconstruct the buoyancy glitch profile.
We show the advantage of buoyancy coordinate in Sec 2.
We derive the theoretical expressions that link the Fourier spectra of 
period spacings to the derivatives of Brunt glitch profile in Sec 3.
These expressions are verified with MESA stellar models in Sec 4.
Before applying to observed g modes in SPB and $\gamma$ Dor stars in Sec 7,
we discuss the effects of rotation, overshooting and mixing on the period spacing
 in Sec. 6 and Sec 5, respectively. We conclude in Sec 8.

\section{Gravity mode period spacing and Buoyancy coordinate}

In fluids with density stratification, a parcel displaced vertically oscillates at
the Brunt-V\"ais\"al\"a (buoyancy) frequency $N= \sqrt{-(g/\rho_0)d\rho/dz}$.
In regions with $N^2 >0$ where gravity wave propagates, buoyancy acts as the restoring
force of the oscillation. Conversely, in convectively unstable regions with $N^2 <0$,
gravity modes are evanescent.

\FloatBarrier 
\begin{strip}
   \centering
 \includegraphics[width=\textwidth,angle=0]{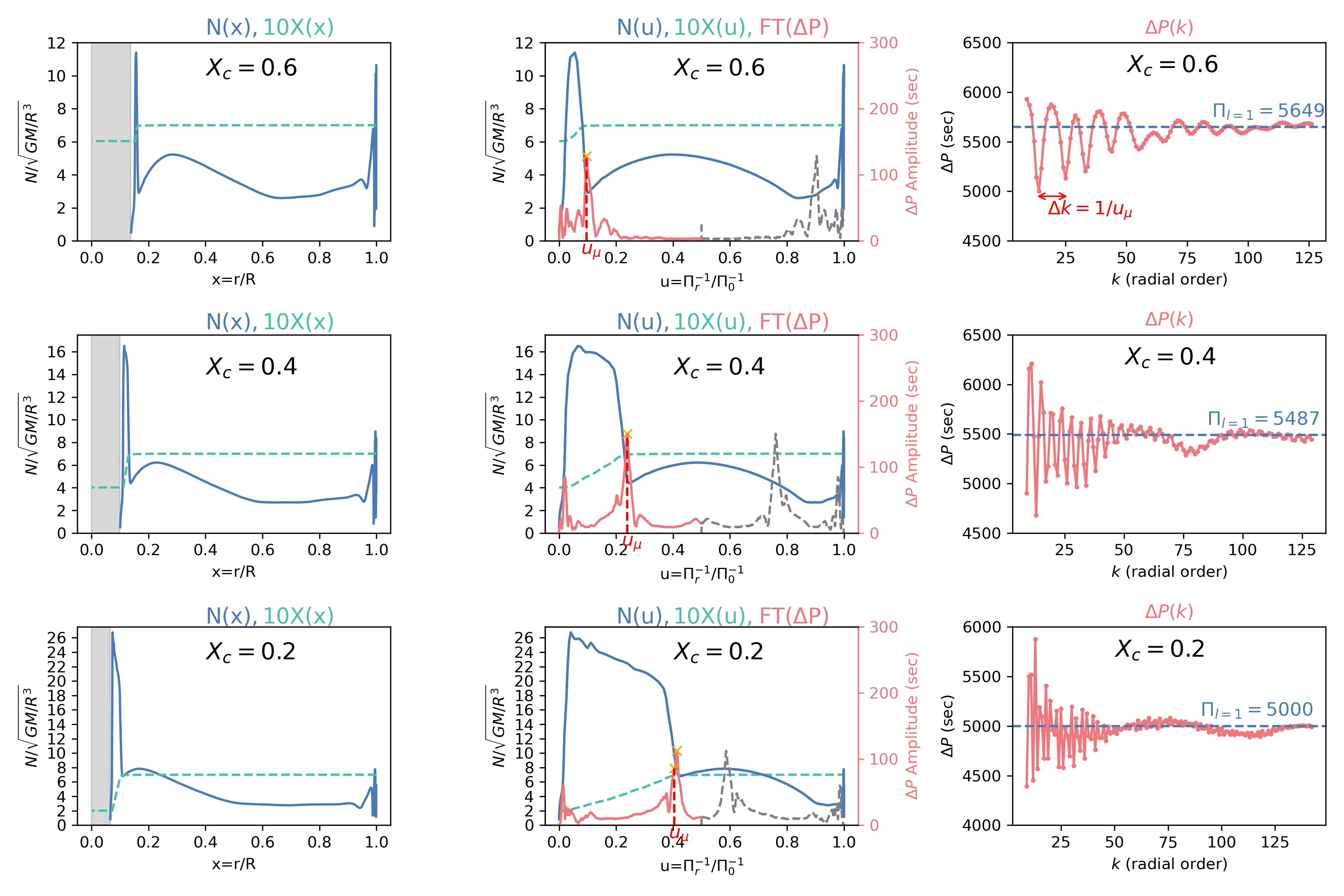}
 {\captionsetup{hypcap=false}
 \captionof{figure}{Brunt--V\"ais\"al\"a (buoyancy) profiles in the left and middle columns are plotted
  as a function of radial coordinate $x=r/R$ (left) and the buoyancy coordinate $u$ (middle), 
   for an $M=3.2M_{\odot}$ MESA stellar model with solar metallicity and convective overshooting $f_{ov}=0.015$.
   The hydrogen mass fraction $X$ is shown by green dashed lines (multiplied by 10 for clarity).
   The gray-shaded regions denote the receding convective core.
   The dipole gravity-mode period spacings ($\Delta P_k$) are shown in the right panel,
    and their Fourier amplitude spectra $FT(\Delta P)$ are overplotted in the middle panels (red lines + super-Nyquist gray parts).
   The dominant Fourier spectra peak aligns with the sharp drop points of the Brunt profile and is labeled as $u_{\mu}$ in red.}
   \label{fig1} }%
\end{strip}

   \FloatBarrier 

 Figure 1 (left column) presents the Brunt profile $N$ (in units of $G = M = R = 1$) as a function of
  radial coordinate $x=r/R$ for an $M=3.2M_{\odot}$ main-sequence stellar structure model from MESA
 \citep{Pax11,Pax13}.
The `gs98' chemical composition and solar metallicity are adopted.
Ledoux criterion is employed for convection instability and we use
exponential convective core overshooting \citep{Her00}.

As the star evolves on the main sequence with central hydrogen
depleting from $X_c=0.6$ to $X_c=0.2$, the convective core (where $N \approx 0$, indicated by the gray-shaded regions 
to the left of the Brunt profile $N$) recedes and leaves a 
chemical-composition gradient region. Green dashed lines show the hydrogen mass fraction $X$.

It is convenient to use the buoyancy coordinate $u$ (the normalized buoyancy radius),
rather than the radial coordinate $x=r/R$, since g modes are naturally described by their 
buoyancy travel time. In this coordinate, nodes of high-order g-modes end up evenly spaced,
 with $u$ defined as:
\begin{equation}
   u(r)=\Pi^{-1}_r/\Pi^{-1}_0 =\frac{\int^x_{x_1}{N(x')/x' dx'}} {  \int^1_{x_1}{N(x')/x' dx'}}
\end{equation}
where, $x_1$ is the inner boundary of the g-mode cavity.
And the two differential elements $du$ and $dx$ are related by
\begin{equation}
   \Pi^{-1}_0 du=  \frac{N}{x}dx \propto \frac{dx}{v_{gr} },   
\end{equation} 

This shows that $du$ is proportional to the buoyancy travel time across $dx$, with the gravity-wave 
group velocity being $v_{gr}=\frac{\omega^2/ \sqrt{l(l+1)}} {N/r}$.

Using $u$ coordinate "stretches" the the deep interior of the star and
"compresses" the outer parts so that the g-mode cavity looks like an almost constant
slab of unit length, making structural diagnostics and glitches easier to spot.

Figure 1 (middle column) shows the Brunt profile in the $u$ coordinate
for SPB star models at different main-sequence stages $X_c=0.6, 0.4, 0.2$. It is easier to see that
as the star evolves, the Brunt-profile bump shifts to the right 
 and its amplitude increases.

In regions where $N/r$ is large (e.g. near a sharp composition gradient at 
the convective-core boundary), a small change in $x$ 
produces a large jump in $u$. These regions act as glitches to gravity waves,
which induce a variation in the g-mode period spacings and mode nodes also
 cluster in high-$N$ zones (mode trapping), especially in compact pulsators (sdB stars and white dwarfs).

With a smooth Brunt profile, the pulsations of gravity modes observed 
in $\gamma$ Dor and SPB stars are equally spaced in periods \citep{Tas80}.
The asymptotic period spacing is $\Pi_l=\Pi_0/ \sqrt{l(l+1)}$, where $\Pi_0=2 \pi^2 (\int_{N^2>0}(N/r)dr)^{-1}$
,which measures the "buoyancy travel time" across the g-mode cavity $N^2>0$. 

Thus the pulsation periods are well approximated by the linear relation \citep{Van16}:
\begin{equation}
   P_k \approx  \frac{\Pi_0}{\sqrt{l(l+1)}}(k + k_0) = a(k+k_0)
\end{equation}
with $k$ being the radial order of the mode, and $k_0$ being a constant close to $0.5$.
$k_0$ may also absorb an additional constant due to phase offset and mode mis-identification of the radial order $k$.
For Fourier analysis, $k_0$ can be conveniently adjusted by an additive constant.
When roation is included in the Traditional Approximation, 
the expression remains unchanged except for
 $\sqrt{l(l+1)} \rightarrow \lambda_{l,m,s}$ (see Sec. 6).

Figure 1 (right column) confirms that the equally spaced periods of g modes 
in the SPB model, with $l=1$ and $m=0$, oscillating around the asymptotic 
period spacing $\Pi_l$ (blue dashed lines).

In reality, stellar interiors often contain sharp features
relative to the wavelength of the gravity waves in question.
This is particularly true for the convective-core boundary and 
composition-transition points, where the chemical abundances
and hence the Brunt profile $N$ undergoes abrupt changes. In the left and middle columns
 of Figure 1, the hydrogen mass fraction $X$ (green dashed lines) shows 
 sharp drops at the chemical composition transition region, 
 which is the main source of the Brunt profile changes.

It is well known that the sharp drop in the Brunt profile $N$, produces 
a periodic variation in the g-mode period spacings $\Delta P$.
In particular, the variation period in $\Delta P$ counted in radial order $k$ 
and denoted $\Delta k$, is exactly the reciprocal buoyancy radius 
at the chemical-composition glitch $u_{\mu}$ \citep{Mig08, Wu18},

\begin{equation}
   \Delta k=\frac{1}{u_{\mu}}.
\end{equation}

The peak of the Fourier spectrum of the period spacings $FT(\Delta P_k)$
(red lines in the middle column of Figure 1) aligns with $u_{\mu}$ where 
the Brunt profile $N$ drops sharply.
For the model with $X_c=0.6$, $u_{\mu}$ is about $0.1$ as seen in the middle 
panel, implying a variation period in $\Delta P$ of about 10 radial orders
 ($\Delta k \approx 10$). All the Fourier spectra are calculated with radial 
 orders from $9$ to $125$. Note that since the Nyquist frequency is $0.5$, 
 the Fourier spectra show a super-Nyquist reflection around this value.

The receding convective core leaves behind a chemical composition gradient region,
as indicated by the increasing segments of the green lines, which gives rise to
the expanding Brunt profile bump $N$ (middle column of Figure 1).
As the star evolves from $X_c=0.6$ to $X_c=0.2$, the Brunt profile bump
widens and $u_{\mu}$ shifts to larger values, producing higher and 
higher frequency variations in $\Delta P$ (Figure 1 right panel). The variation periods change from $\Delta k \approx 10$ to $\approx 4$, and $\approx 2.5$ for $X_c = 0.6, 0.4$ and $0.2$, respectively. Note that for $X_c=0.2$, the dominant variation frequency is very high so that consecutive peaks/troughs are very close. Visually, the subdominant, low-frequency variation can be seen more easily, with the dip at $k \approx 110$.

Concurrently, this increases the integral of $N/r$, and reducing 
the asymptotic period spacing $\Pi_l$ (from
 $5649$ sec at $X_c=0.6$ to $5000$ sec at $X_c=0.2$).

The sharp-increasing part of Brunt profile $N$ also generates
a peak in $FT(\Delta P_k)$ near $u=0$, but this peak typically
has a lower amplitude than the peak at $u_{\mu}$.
This explains the slow-varying trend of the period spacing $\Delta P_k$ observed
in the right panels of Figure 1, particularly in $X_c=0.4$ and $0.2$ models.

In the following sections, we will take a step further and
 show that the Fourier spectra of the period spacings
are directly linked to the Brunt glitch profile and its derivatives.


\section{Connecting Buoyancy Glitch profile derivatives to Fourier spectra of period spacings}

For gravity modes observed in intermediate and high-mass stars,
the Brunt profile $N$ changes sharply at the chemical-composition 
gradient region near the convective-core boundary. Such buoyancy glitches,
 $\delta N$, induce a frequency perturbation $\delta\omega$, which, when 
 written as the relative pulsation-period perturbation 
 at radial order $k$, takes the form of \citep{Mon03, Mig08, Zha23}:

\begin{equation}
   \frac{\delta P_k}{P_k} \approx - \int^1_0{\frac{\delta N(u)}{N(u)}[1+\sin(2\pi (k+k_0) u)]du}
\end{equation}
where $N$ is the Brunt profile, and we write $z(u)$ as $z(u)=\delta N/N$.

The observed pulsation periods of g modes are, 
\begin{equation}
   P_{k,obs}= P_{k} + \delta P_k .
\end{equation}

It is common to represent the period spacing at the $k$-th radial order $\Delta P_{k,obs}$ as
a function of the periods $P_k$. By applying eq. 3, one can readily recast this dependence 
in terms of the radial order $k$:

\begin{equation}
   \Delta P_{k,obs}(k) =P_{k+1}-P_{k} + (\delta P_{k+1} - \delta P_k) \approx a +\frac{d \delta P_k}{dk} \cdot 1
\end{equation}
The above equation indicates that the period spacing $\Delta P_{k,obs}$ 
is approximately a constant $a$, with a small perturbation originating from
$d\delta P_k/dk$.

Starting from eq. 5 and treating the term in the brackets [] as the derivative of $C(u)=u-\cos[2\pi (k+k_0) u]/(2 \pi (k+k_0))$,
we can integrate by parts to obtain:
\begin{equation}
   \delta P_k/P_k =+ \int^1_0{z'(u)[u-\cos[2\pi (k+k_0) u]/(2 \pi (k+k_0))]du}
\end{equation}
the boundary terms vanish because it is natural to assume Brunt glitch profiles 
in stars satisfy $z(u=0)=0$ and $z(u=1)=0$
\footnote{We are not interested in the near-surface region, where $N$ can be large}.
Here, we have denoted $z'(u)=dz/du$. Multiplying by $P_k$ and applying eq. 3, we obtain,

\begin{equation}
   \delta P_k = a \int^1_0{z'(u)[u (k+k_0)-\cos[2\pi (k+k_0) u]/(2 \pi)]du}
\end{equation}

Taking the derivative with respect to $k$ on both sides, we obtain,
\begin{equation}
  d \delta P_k /dk = a \int^1_0{z'(u)[u + \sin[2\pi (k+k_0) u](2\pi u)/(2 \pi)]du}.
\end{equation}

Thus we can rewrite observed period spacing as 

\begin{align}
   \Delta P_{k,obs}\approx  a + a \int^1_0{z'(u)u [1 + \sin[2\pi (k+k_0) u]]du} \\
    =a + a \int^1_0 z'(u)u du + a \int^1_0[ z'(u) u \frac{e^{i 2 \pi (k+k_0) u}-e^{-i 2 \pi (k+k_0) u}}{2i} ]du \\
    =a - a \int^1_0 z(u) du + a \int^1_0[ z'(u) u \frac{e^{i 2 \pi (k+k_0) u}-e^{-i 2 \pi (k+k_0) u}}{2i} ]du \\
\end{align}

The second term is a negligible constant, and we have
\begin{equation}
   f(k)=\frac{\Delta P_{k,obs} - a}{a} \approx \int^1_0 \frac{z'(u) u}{2 i}[ e^{i 2 \pi (k+k_0) u}-e^{-i 2 \pi (k+k_0) u} ]du \\
\end{equation}

Take Fourier transform with respect to $k$ and transform to the "frequency" space $\xi$ on both sides,
\begin{align}
   \mathcal{F} (f(k))(\xi)=\int^{+\infty}_{-\infty} f(k) e^{-i 2 \pi \xi k} dk  \\
 = \int^{+\infty}_{-\infty} \left(\int^1_0 \frac{z'(u) u}{2 i}[ e^{i 2 \pi (k+k_0) u}-e^{-i 2 \pi (k+k_0) u} ]du\right)e^{-i 2 \pi \xi k} dk \\
 = \int^1_0 \frac{z'(u) u}{2 i} \left(e^{i 2 \pi k_0 u}\delta(u-\xi) -e^{-i 2 \pi k_0 u}\delta(u+\xi) \right) du \\
 =\frac{z'(\xi) \xi }{2 i} e^{i 2 \pi k_0 \xi} \  (\rm{for} \  0 \leq \xi  \leq 1) \\
\end{align}



The one-sided Fourier amplitude spectrum (denoted by "FT") of $\Delta P_{k,obs}$ is,
\begin{equation}
 FT(\Delta P_k -\Pi_l)=2 |\mathcal{F}(\Delta P_k -\Pi_l)| = \Pi_l \left |\frac{d z(\xi)}{d\xi}\xi \right | =\Pi_l \left |\frac{d z(\xi)}{d\ln \xi} \right |
\end{equation}
with $a=\Pi_l=\Pi_0/\sqrt{l(l+1)}$.

Since an extra $\xi$ term appears in front of the derivative $dz/d\xi$,
the near-zero peaks in $FT(\Delta P_k)$ are strongly suppressed.
This explains why the low-frequency peak is typically lower than the
dominant one at the sharp drops in the Brunt profile $u_{\mu}$ (e.g. Figure 1 middle panels) .


The same derivation can be applied to the 
relative period perturbation
 $\frac{\delta P_k}{P_k}$, and we obtain,
\begin{equation}
   \mathcal{F} (\frac{\delta P_k}{P_k})(\xi) \approx -\frac{1}{2i} z(\xi) e^{i 2 \pi k_0 \xi}.  
\end{equation}

Thus the one-sided Fourier amplitude spectrum is
\begin{equation}
 FT(\delta P_k/P_k)=2 \left |\mathcal{F}(\frac{\delta P_k}{P_k})\right | = |z(\xi)| =\left |\frac{\delta N}{N}(\xi)\right |
\end{equation}

In practice, $k$ (interpreted as 'time') are integers and $\mathcal{F}(\Delta P_k)$ is the discrete Fourier transform
which transforms $k$ to the $\xi$ ('frequency') space ranging from $-0.5$ to $0.5$, 
with 0.5 being the Nyquist frequency.
$u$ is from $0$ to $1$, and it is convenient to use $u$ also as
 the corresponding frequency space instead of $\xi$ once the $0.5$ offset has 
 been accounted for.

Observationally, the two Fourier spectra are related by,
\begin{equation}\label{eq:ftft}
   \frac{d FT(\delta P_k/P_k)}{d\ln u} = FT(\frac{\Delta P_k-\Pi_l}{\Pi_l}) 
\end{equation}

Since $FT(\Delta P_k)$ and $FT(\delta P/P)$ depend only on the
glitch profile $z=\delta N/N$, there is no dependence on mode spherical degree $l$. 
The observed period spacings for different $l$ should exhibit
variations with the same periodicity.

The expressions derived above are verified in Sec. 4.

Thus, the peaks in the Fourier spectra of 
the period spacings align with the locations 
where $dz/du$ peaks, which typically correspond to sharp drops/dips
 and jumps/kinks in the Brunt profile. The height of the 
 peaks reflects the amplitude of the Brunt glitch derivative.


Once the glitch profile $z(u)=\delta N/N$ is obtained, 
we can further recover the Brunt profile in physical
 coordinates, $x=r/R$. Rearranging equation 2, we obtain,
\begin{equation}\label{eq:dudx}
   dx/x=\Pi^{-1}_0 du/N(u).   
\end{equation}
Integrating from the surface ($u=1,x=1$) down to an arbitrary u, we obtain:
\begin{equation}
   \ln x(u)=\Pi^{-1}_0 \int^u_1{\frac{d(u')}{N(u')}}. 
\end{equation}


\section[Verification of FT(Delta P) relations with MESA models]{Verification of $FT(\Delta P)$ relations with MESA models}

\subsection{Verification with Brunt profiles and g modes in SPB models}

Based on a mid-main-sequence ($X_c=0.4$) MESA model with $M=4.5M_{\odot}$
and the corresponding dipole ($l=1$) gravity modes calculated with the GYRE oscillation code
 \citep{Tow13,Tow18,Sun23}, we verify the theoretical relations derived in the previous section.

Figure 2(a) shows the Brunt profile $N$ as a function of 
buoyancy coordinate $u$. The sharp jump/drop features in $N$ act
as glitches to the gravity waves, which induce a periodic
 variation in the g-mode period spacings. We can decompose
 the Brunt profile to a smoothed part $N$ and a glitch part $\delta N$ 
 (exaggerated for clarity).
We also show the glitch $\delta N/N=z(u)$, the relative perturbation
 of the Brunt profile. It varies around zero with the same order 
 as $\delta P/P$ (Figure 2(g)). The glitch $\delta N$ has 
 been scaled for clarity in Figure 2(a).

It is important to note that we distinguish between $\delta P$, the deviation of pulsation period
 from the linear, asymptotic relation  ($P_k \propto k$) as shown 
in Figure 2(b), and the observed period spacing $\Delta P_k =P_{k+1}-P_k$
 (often used in the literature\footnote{Some literature uses $\delta P$ as the period spacing as well.}).
Figure 2(d) illustrates that $\Delta P$ oscillates around a mean value
 $\Pi_l$. The normalized period spacing $(\Delta P_k-\Pi_l)/\Pi_l$ 
is shown in Figure 2(e), which oscillates around zero with an order
of $\approx 0.1$. Similarly, the relative period perturbation $\delta P/P$
is displayed in Figure 2(h). Note that $\delta P/P$ is much smaller, only
on the order of about $\lesssim 1\%$.

Figure 2(f) compares the one-sided Fourier amplitude 
spectrum $FT(\Delta P_k -\Pi_l)/\Pi_l$ with the derivative
of Brunt glitch profile $|d z(u)/d\ln u|$. This verifies eq. 21.
Although in practice, the $FT(\Delta P_k -\Pi_l)/\Pi_l$ is much smoother,
due to its limited resolution.

Figure 2(i) highlights the comparison of $FT(\delta P/P)$ and
$|z(u)|=|\delta N/N|$. This validates eq. 23.
Again, the Fourier spectrum represents a 'smoothed 
version' of the true glitch profile $z(u)$.

Finally, the two Fourier spectra $FT(\delta P/P)$ 
and $FT(\Delta P_k -\Pi_l)/\Pi_l$ are directly
linked by eq.~\eqref{eq:ftft}. And the comparison of red and green lines (respectively)
in Figure 2(c) supports this relation. Note that the 
Fourier spectra show symmetric super-Nyquist peaks,
it seems that the peak at $0.7$ agrees better with 
$u dFT(\delta P/P)/du$ than the real peak at $0.3$.
This can be explained by a phase offset of $0.5$ between
$u$ and $\xi$ (see Sec. 3).
\FloatBarrier 
\begin{strip}
   \centering
 \includegraphics[width=\textwidth,angle=0]{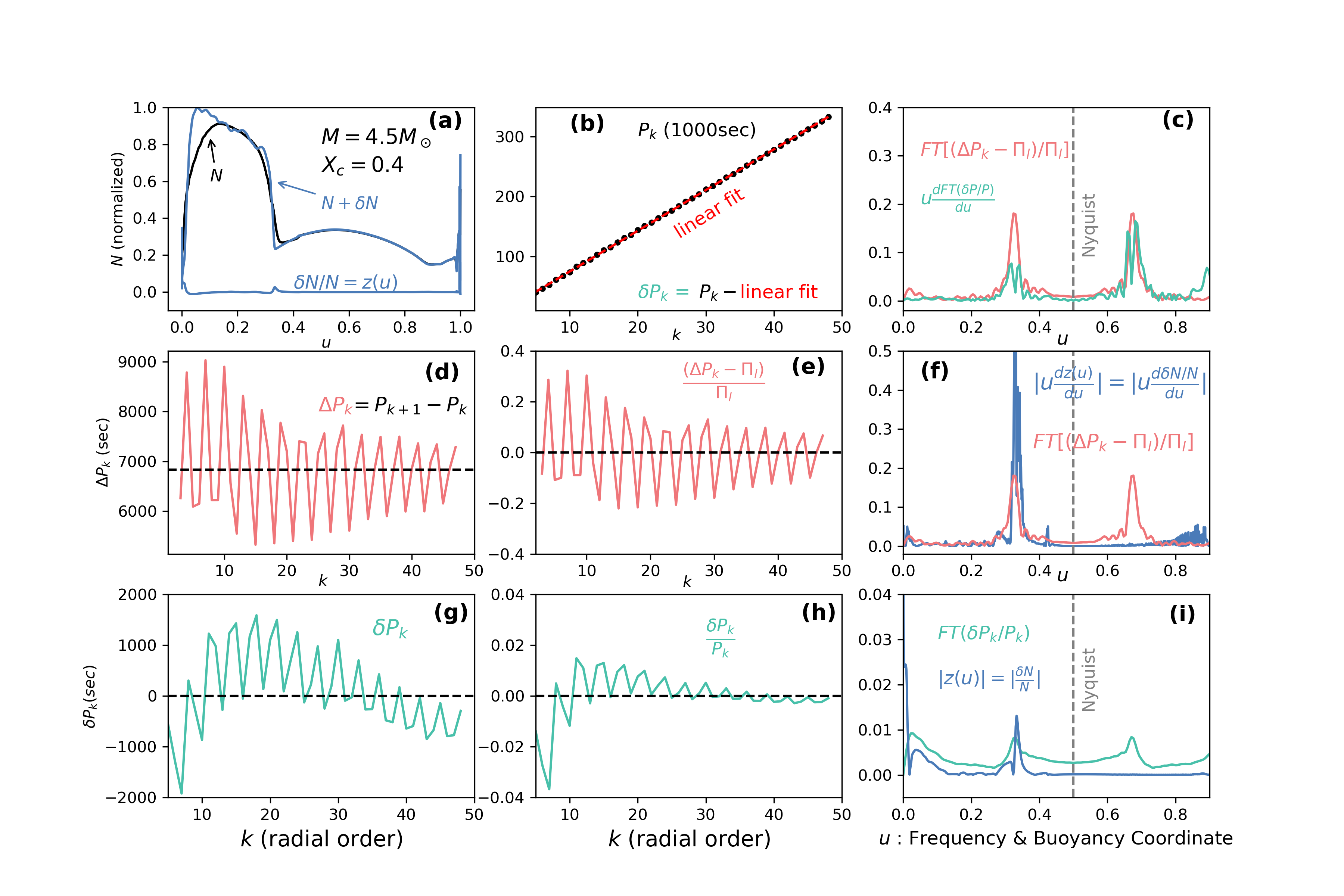}
 {\captionsetup{hypcap=false}
 \captionof{figure}{\textbf{(a)}: Brunt profile decomposed to a smooth part $N(u)$ and a glitch part $\delta N(u)$. \textbf{(b)}: Pulsation periods vs radial order $k$ ($l=1$ g modes). \textbf{(c)}: Comparison between $FT(\Delta P)$ and the derivative of $FT(\delta P/P)$. \textbf{(d)}: Period spacings $\Delta P_k$. \textbf{(e)}: relative period spacings $\Delta P_k$ with respect to the asymptotic value $\Pi_l$. \textbf{(g)}: ($\delta P_k$) Deviations (perturbations) of pulsation period from asymptotic values. \textbf{(h)}: relative perturbations of pulsation periods ($\delta P_k/P_k$). \textbf{(f)}: Comparison of the Fourier spectrum of $\Delta P_k$ and the buoyancy glitch derivatives. \textbf{(i)}: Comparison of the Fourier spectrum of $\delta P_k/P_k$ and the buoyancy glitch profile $\delta N/N$. All the plots are based on a middle-main-sequence ($X_c=0.4$), $M=4.5M_{\odot}$ stellar model and its dipole g modes.     }
   \label{fig2} }%
\end{strip}

   \FloatBarrier 


\section{Effects of convective overshooting and mixing} 

We calculate stellar models with the mass $M=1.6M_{\odot}$ (representing a gDor star)
 and $M=3.2, 4.5, 6.0M_{\odot}$ (spanning roughly the SPB star mass range).
We adopt the convective-core overshooting parameter $f_{ov}=0.015, 0.025$,
and the envelope mixing parameter $D_{min}=10,20,50,100$.

The right panel of Figure 3 shows the running
of period-spacing varying frequency ($u_{\mu}$), i.e., the peak
frequency of the normalized $\Delta P_k$ Fourier spectra, along with the
central hydrogen mass fraction $X_c$ (a proxy for main-sequence age).
A strong, near-linear correlation can be observed, with the best-fitting
linear relation given by:

\begin{equation}
u_{\mu} = (X_c -0.70)/(-1.13).
\end{equation}

To be more specific, $\gamma$ Dor models with $M=1.6M_{\odot}$
all have $u_\mu$ values lower than $0.5$. For SPB stars,
 most main-sequence models with $X_c \gtrsim  0.1$ have 
peak frequencies $u_\mu$ smaller than $0.5$.
However, for B-star models that are close to TAMS,
 the true peak in $FT(\Delta P_k)$ can exceed 0.5, 
 which reflects back to the sub-Nyquist region and produces
 a spurious peak symmetric to the real one. 

Figure 1 and 3 (right column) can be compared with \citet{Mig08}.
In their Fig 15, they show an $M=1.6M_{\odot}$ $\gamma$ Dor model,
for which $X_c=0.5,0.3,0.1$ correspond to
$u_\mu = 0.09, 0.16, 0.33$, respectively.
For the SPB star model with $M=6.0M_{\odot}$ at the same $X_c$, the 
corresponding $u_\mu$ values are larger: $0.15,0.32,0.50$ (their Fig 16).

For our calculations, the upper-right panel of Figure 3 shows that 
at the same main-sequence age ($X_c$), models with larger overshooting result in
higher values of $u_{\mu}$. And this effect is more pronounced for the $1.6M_{\odot}$ model 
than the SPB star models. This agrees with Miglio's results 
with convective-core overshooting. For SPB stars, $u_{\mu,ov}$ is only slightly
larger than the non-overshooting counterpart $u_{\mu}$. Whereas for the gDor model, the difference
is much more significant. As shown in their Fig 17 and Fig 18 for 
the $X_c=0.3$ model with $\alpha_{ov}=0.2 H_p$, the difference in $u_{\mu}$
is approximately $0.05$.

Even with different levels of overshooting,
the tight relation between $u_{\mu}$ and $X_c$ still holds,
with a scatter less than $0.1$ in $X_c$.
Similarly, the lower-right panel illustrates the effect of 
envelope mixing $D_{min}$. At the same $X_c$, a larger $D_{min}$
 leads to a slightly lower $u_{\mu}$, but the scatter is somewhat 
smaller than that in the overshooting case.

Another observation is that, within the SPB star mass range,
this tight correlation exhibits only a very weak
 dependence on stellar mass, indicating that $u_{\mu}$ is
a good age probe for any SPB stars regardless of their masses. 
For the $\gamma$ Dor star model, the $u_{\mu}-X_c$ curve
lies well below the SPB-star fitting line.

We emphasize here that even with different overshooting and
 mixing parameters, the tight relation between $u_{\mu}$
  and $X_c$ still holds. The $FT(\Delta P_k)$ peak frequency
$u_{\mu}$ can be used as an excellent age indicator for SPB stars,
and still a reliable age probe for $\gamma$ Dor stars.

Miglio et al. also show that element diffusion smooths out the Brunt profile, leading to smaller
 local $dN/du$ and the peak in $FT(\Delta P_k)$ becomes lower.
Consequently, diffusion reduces the visibility of variations in $\Delta P$ (their Fig 21, 22).

Based on MESA models with different envelope mixing parameters $D_{min}=10,20,50,100$
and stellar masses $M=3.2, 4.5, 6.0M_{\odot}$, we measure 
the peak amplitude of the corresponding $l=1$ g-mode period-spacing Fourier spectra and plot the
result in the lower-middle panel of Figure 3.
As expected, the Brunt profile derivatives and therefore the 
amplitude of $FT(\Delta P)$ depend sensitively on the
mixing parameters, with larger $D_{min}$ resulting in significantly smaller
amplitude.

\section[The effect of rotation on FT(Delta Pk)]%
        {The effect of rotation on $FT(\Delta P_k)$}
        
The derivations in Section 2 can be extended to rotating stars.
Under the traditional approximation for rotation, the pulsation periods of
gravity modes in the co-rotating frame can be described by,

\begin{equation}
   P_{k,co} = \frac{\Pi_0}{\sqrt{\lambda_{l,m,s(k)}}}(k + k_0)
\end{equation}

We still have the variational relation for pulsation period perturbations,
\begin{equation}
   \frac{\delta P_{k,co}}{P_{k,co}} \approx \int^1_0{z'(u)[u-\cos(2\pi (k+k_0) u)/(2\pi(k+k_0))]du}.
\end{equation}

Equation 7 can be rewritten as:

\begin{equation}
\begin{aligned}
\Delta P_{k,\mathrm{co},\mathrm{obs}}(k)
&= P_{k+1,\mathrm{co}}-P_{k,\mathrm{co}}
   + \bigl(\delta P_{k+1,\mathrm{co}} - \delta P_{k,\mathrm{co}}\bigr) \\
&\approx \Delta P_{k,\mathrm{co}} + \frac{d\,\delta P_{k,\mathrm{co}}}{dk}
\end{aligned}
\end{equation}

Following Bouabid et al.\ (2013), the asymptotic, co-rotating frame period spacing
is given by, 
\begin{equation}
   \Delta P_{k,co} \approx \frac{\Pi_0}{\sqrt{\lambda_{l,m,s(k)}}}\frac{1}{(1+0.5\frac{d\ln\lambda_{l,m,s(k)}}{ds})} \approx \frac{\Pi_0}{\sqrt{\lambda_{l,m,s(k)}}}
\end{equation}
where $\lambda_{l,m,s(k)}$ is the eigenvalue of the Laplace tidal operator,
generally a slowly varying function of $k$.

\begin{align}
   \frac{d\delta P_{k,co}}{dk} \approx \frac{\Pi_0}{\sqrt{\lambda_{l,m,s(k)}}} \int^1_0{z'(u)[u+u\sin(2\pi (k+k_0) u)]du} + \\
   \Pi_0 \int^1_0{z'(u)[u(k+k_0)-\cos(2\pi (k+k_0) u)/(2\pi)]du}\frac{d}{dk}(\frac{1}{\sqrt{\lambda_{l,m,s(k)}}})
\end{align}

We can ignore the second term again using slowly-varying assumption of $\lambda$ with respect to $k$.

Define $\tilde{\Pi}_l=\Pi_0/\sqrt{\lambda_{l,m,s(k)}}$,
we obtain a similar result as eq 11,

\begin{equation}
   \frac{\Delta P_{k,co,obs}- \tilde{\Pi}_l }{\tilde{\Pi}_l} \approx  \int^1_0{z'(u)u [1 + \sin[2\pi (k+k_0) u]]du}.
\end{equation}

To a good approximation, the results on the Fourier spectra relations in Sec. 2
still hold, with the only difference being that
the asymptotic period spacing $\Pi_l$ is replaced
 by $\tilde{\Pi}_l=\Pi_0/\sqrt{\lambda_{l,m,s(k)}}$,
 and we work in the co-rotating frame periods $P_{k,co}$ 
 and their spacings $\Delta P_{k,co}$.
We can then apply Equation 21 and 23 to observed
 g modes in rotating stars.

In practical applications, we transform the inertial (observer's) frame 
periods $P_{k,inert}$ to the co-rotating frame 
periods $P_{k,co}$ using
\begin{equation}
   P_{k,co} = P_{k,inert}/\left (1+m \frac{P_{k,iner}}{P_{rot}}\right ),
\end{equation}
where we have used the convention that $m=-1, +1$ 
correspond to prograde and retrograde modes, respectively.

The corresponding period spacings are transformed as follows:
\begin{align}
   \Delta P_{k,co} \approx \Delta P_{k,inert} \left (1-mP_{k,co}/P_{rot} \right )^2 \\
   =  \Delta P_{k,inert}\left (1-\frac{1}{1+P_{rot}/(m P_{k,inert})} \right )^2.
\end{align}

In the next section, we apply the above results to the g modes in the 
rotating SPB star KIC 7760680. 

\section{Applications to observations in g-mode pulsators}

We apply the above results to the observed g modes in two slowly-rotating stars,
 KIC 11145123 and KIC 10526294,
for which the rotational effects on g modes are negligible.

We then apply the rotation-corrected results in Section 6 
to the rotating SPB star KIC 7760680.

\subsection[Application to observations: slowly rotating SPB and gamma Dor stars]%
           {Application to observations: slowly rotating SPB and $\gamma$ Dor stars}

KIC 10526294 (B8V, $\approx 3.2M_{\odot}$,$T_{\rm eff}\approx 11550K$, $\log g \sim 4.1$,
rotation period $P_{rot} \approx 190$ days) is a SPB pulsator that has been studied in detail by \citet{Pap14} and
 \citet{Mor15}. The $l=1$ prograde g modes have been 
 identified and the period spacing pattern is plotted in 
 the upper-middle panel of Figure 3.   

Figure 3 (upper left panel) shows $FT(\Delta P)$,
and the dominant peak is at $u_{\mu} \approx 0.059$,
which translates to $X_c=0.63$ using eq. 27. This is
consistent with $X_c=0.627$ from the detailed seismic modeling
 of \citet{Mor15}.
The peak amplitude of $FT(\Delta P)$ is 230.4 seconds, and is also 
shown in the lower-middle panel of Figure 3.

KIC 11145123 ($M \sim 1.46M_{\odot}$) also rotates very slowly; the inferred rotation period 
is about $100$ days \citep{Kur14}. The period spacing agrees with
a stellar model close to terminal-age-main sequence (TAMS), and the 
estimated $X_c$ is as low as $\approx 0.033$ although with large uncertainty. 
Our Fourier analysis of the period spacing reveals a peak at $u_{\mu} \approx 0.425$ (Figure 3, lower left),
 which is consistent with the result of low $X_c$ when comparing with 
the $1.6M_{\odot}$ model as shown in the upper-right panel of Figure 3 (purple lines).
It appears that a slightly lower mass ($< 1.6M_{\odot}$) is more appropriate,
similar to the value adopted in \citet{Kur14}.
More detailed seismic modeling has been carried out by 
\citet{Hat21}, who found that non-standard modeling is required,
supporting the argument that KIC 11145123 is likely a blue straggler.

We find that for a typical $1.6M_{\odot}$ model, the highest peaks of $FT(\Delta P)$
usually correspond to the sharp-rising side of the Brunt profile (convective-core boundary),
 which is at very low $u$. The downward side of the Brunt profile is smoother than SPB star models,
and produces a broad, subdominant peak. Thus in general, applying the $FT(\Delta P)$
technique to $\gamma$ Dor stars is more challenging than SPB stars.

\begin{figure*} 
      \centering
    \includegraphics[width=\textwidth,angle=0]{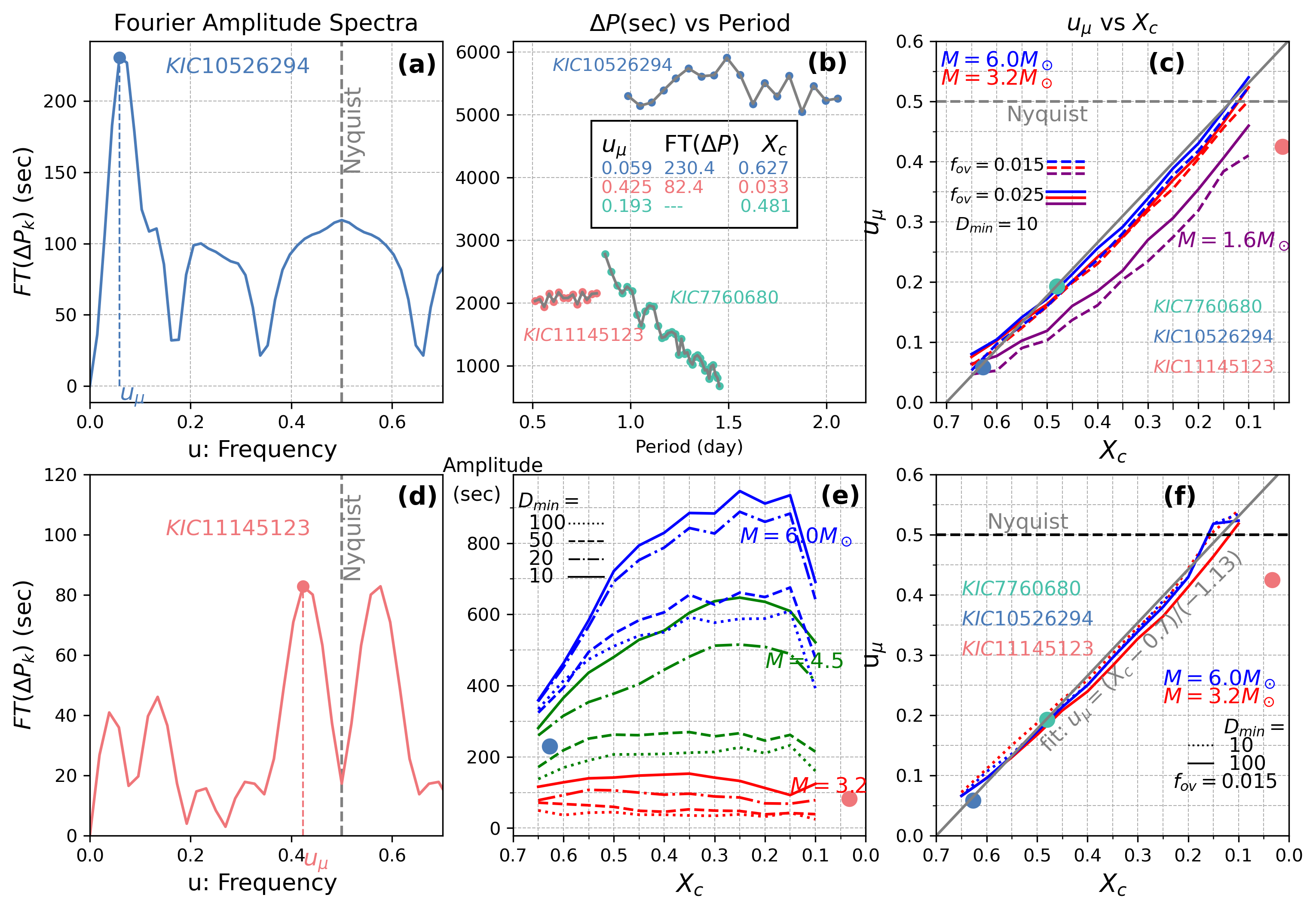}
    \captionof{figure}{Panel \textbf{(b)}: Observed g-mode period spacings $\Delta P$ and periods for KIC 10526294, KIC 11145123, KIC 7760680. The Fourier spectra of the former two stars are shown in panel \textbf{(a)} and \textbf{(d)}, respectively. The dominant peaks are labeled by the filled circles with corresponding frequency $u_{\mu}$. Panel \textbf{(c)} and \textbf{(f)} presents the g-mode $\Delta P$ variational frequency $u_{\mu}$ as a function of central hydrogen mass fraction $X_c$ for MESA models of masses $M=6.0$ (blue), $3.2$ (red) and $1.6M_{\odot}$ (purple). Panel \textbf{(c) } shows models with different convective-core overshooting $f_{ov}$ and \textbf{(f)} displays the effect of the envelope mixing $D_{min}$. Panel \textbf{(e)} demonstrates the amplitudes of FT($\Delta P$) for MESA models with different masses and $D_{min}$.  }
      \label{fig3}%
\end{figure*}
   

\subsection{Application to the rotating SPB star KIC 7760680}

KIC 7760680 is a rotating ($f_{rot}\approx 0.48 $day$^{-1}$) SPB star.
\citet{pap15} identified 36 $l=1$ prograde g modes from Kepler light curves.
 Spectroscopic observations placed the star at the low-mass end of the SPB instability strip.
This agrees with the further seismic modeling by \citet{Mor16}, who derived the best-fitting model with a mass of 
 $M \approx 3.25$ and central hydrogen fraction $X_c=0.481$.

In the left two panels of Figure 4, we show the observed $l=1,m=-1$ (prograde) g-mode period 
spacings in both inertial frame $\Delta P_{iner}$ (green) and co-rotating frame
 $\Delta P_{co}$ (blue). Transforming to the co-rotating frame
 removes the downward trend and makes the data ready for Fourier analysis.
The Fourier spectrum of the normalized $\Delta P_{co}$ vs $k$ is shown in the upper-right
panel. The dominant peak is at $u_{\mu} \approx 0.193$, 
indicating a middle-main-sequence model with $X_c \approx 0.48$.
This value is in excellent agreement with the result of 
\citet{Mor16}. In fact,when placed on the  $u_{\mu} -X_c$ 
diagram, KIC 7760680 (green circle) lies directly on the fitting line 
shown in Figure 3 (right two panels).

The Fourier spectrum of the relative period perturbation 
$\delta P_{co}/P_{co}$ is shown in the lower-right panel Figure 4,
which indicates a relative buoyancy glitch $\delta N/N$ of order $0.2\%$ located
at three distinct regions responsible for the periodic variations in 
the g-mode period spacings.
The derivative of this spectrum $d FT(\delta P_{co}/P_{co})/d\ln u$
is calculated and presented in the upper-left panel (red line).
According to eq.~\eqref{eq:ftft}, this is directly equal to 
the Fourier spectrum of the period spacing $FT(\Delta P_k)$ 
(up to a $0.5$ phase offset in practice). It can be seen that the red line
follows the blue line reasonably well, further 
supporting our theoretical results using real observational data.

We did not show the $FT(\Delta P)$ amplitude measurement of 
KIC 7760680 in the lower-middle plot of Figure 3 since the $\Delta P$
amplitude depends on the reference frame and the rotation rate (also illustrated in Figure 4, left column).


\begin{figure*} 
      \centering
    \includegraphics[width=\textwidth,angle=0]{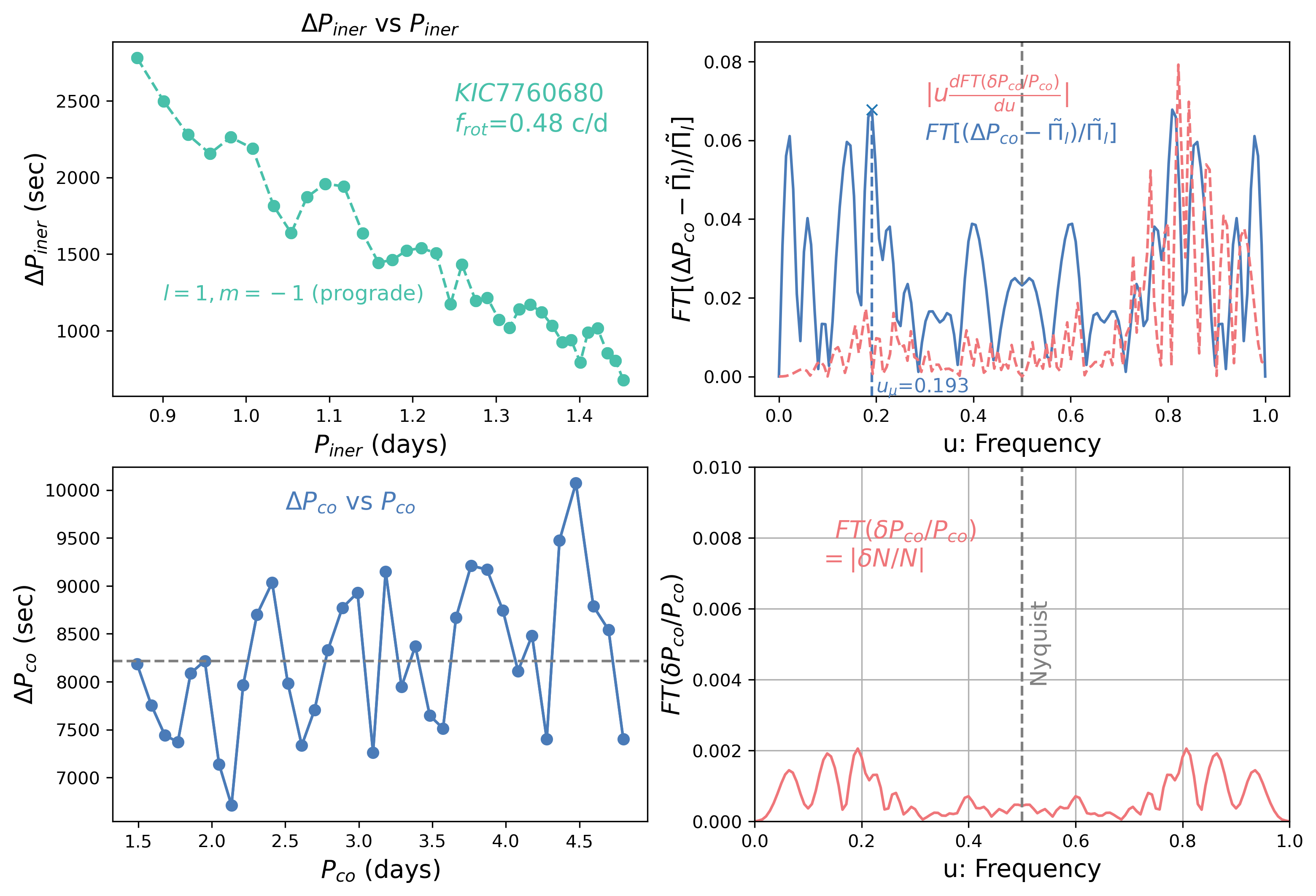}
    \captionof{figure}{\textbf{Left column:} Gravity-mode period spacing
     of KIC7760680 in the inertial frame
      ($\Delta P_{iner}$ vs $P_{iner}$, upper)
      and co-rotating frame ($\Delta P_{co}$ vs $P_{co}$,
      lower panel). \textbf{Right column}: Fourier spectra of normalized $\Delta P_{co}$ (upper) and
      relative period perturbation $\delta P_{co}/P_{co}$ (lower). The two spectra are linked by a relation
       $FT[(\Delta P -\tilde{\Pi}_l)/\tilde{\Pi}_l]=d FT(\delta P/P)/d\ln u$, and compared in the upper panel. }
      \label{fig4}%
   \end{figure*}
\FloatBarrier 


\section{Discussion and conclusions}

\textbf{(1)} Ensemble asteroseismology of g-mode pulsators with $FT(\Delta P)$

Thanks to the straightforward nature 
of this method, conducting ensemble seismology 
on g-mode pulsators becomes much more accessible.
About 611 $\gamma$ Dor stars have been studied in
 \citep{Li19,Li20} and a significant fraction
 show period-spacing variations. \citep{Ped18,Ped21}
 performed detailed seismic modeling of 26 SPB stars
By analyzing the measured $FT(\Delta P)$ frequencies
and their amplitudes, we can uncover correlations 
among stellar age, convective overshooting, mixing parameters, 
rotation, and other related factors. 

Since the Fourier spectra of period spacings and period perturbations
depend only on the Brunt glitch profile $z(u)$ and its derivatives,
there is essentially no l-dependence. This implies that
 $\Delta P$ for different $l$ should show the same variation period. 
We have already verified this in observations and will present the result
in a separate paper. In fact, doubling the number of data points ($l=1$ and $l=2$ modes) can 
further improve the resolution of the Fourier spectra
and thus further constrain the main-sequence age $X_c$ and
 mixing/overshooting parameters. 

 \textbf{(2)} Other g-mode pulsators: compact pulsators, He-burning red 
 giants and beyond

This technique can be naturally applied to other pulsators with g modes.
Compact pulsators such as sdB stars and white dwarfs have several 
different chemical gradient layers/interfaces. Gravity modes can also be trapped in these layers
 \citep{Bra92}, leading to deep dips in the period spacings. 
 By connecting eigenfunctions and using continuity conditions, 
\citet{Cha00} derived analytical expressions for the period spacing
between consecutive trapped modes and related them to the H/He transition layers.
It remains to be seen whether the $FT(\Delta P)$ technique can be applied
to these pulsators with very deep period-spacing dips.

The related He-core-burning red giants also have multiple chemical
 gradient layers. \citet{Bos15} showed that the He-burning shell
 at $u \approx 0.7$ as a glitch, is too smooth to give rise to 
 significant deviations from the asymptotic period spacings 
 However, another glitch at $u \approx 0.17$ due to the discontinuity
in local opacity, induces the periodicity of $\Delta k \approx 6$ in 
the g-mode period spacings. \citet{Mat25} investigated 
the influence of density-discontinuity-induced buoyancy
structural glitches on the period spacings of these oscillation modes
in these stars.

As shown in Figure 4, KIC 7760680 has already shown additional peaks
in $FT(\Delta P)$ at $u \approx 0.02$.This may correspond to 
the low-frequency, subdominant peak close to $u=0$ as shown in
 Figure 1. Detecting multiple periods in $\Delta P$ for more g-mode
pulsators is very promising.

We also verified the $FT(\Delta P)$ technique for post-mass transfer 
stellar models \citep{Wag24, Hen24,Mis25}. These models may have subtle differences
in the Brunt profile due to mass accretion, and the additional
periodic signal can be detected directly from the period spacing
 variations (Wu, Guo \& Li, submitted).

\textbf{(3)} Comparison with classical structure inversions

The technique provides an efficient way to reconstruct the Brunt glitch profile $z(u)$
which can be cross checked with the classical stellar structure inversion
techniques such as OLA and RLS. Moreover, \citet{Van23} explored the applications of 
these inversions for the Brunt profile and density/sound speed pairs.

\textbf{(4)} Linking g-mode asteroseismology to tidal evolution studies

Remarkably, it is interesting to see that the key Brunt glitch profile derivative
 $\frac{d\delta N/N}{d\ln u}$ is extremely similar to the term  $\frac{dN^2}{d\ln r}$
 which appears in the tidal torque expression and is directly linked 
 to the tidal evolution/dissipation timescale and $E_2$.
Its value at the convective-radiative boundary appears in the 
tidal torque expression, which crucially determines the tidal evolution/dissipation timescale 
associated $E_2$ \citep{Zah70, Zah75,Bar20}. Indeed, internal gravity waves (IGW)
are generated at the convective-radiative boundary and propagate outward in intermediate/massive stars 
or inward in low-mass stars such as the Sun. Tidal Love number $k_{lm}$, especially $k_2$ also depends
on this derivative, which is key to the apsidal motion of binary stars.
Linking the observed g-mode period spacings to the derivative of the Brunt profile
 (glitches) opens an avenue to connect gravity-mode 
 asteroseismology to the tidal evolution studies.

 \textbf{(5)} Limitations of the $FT(\Delta P)$ technique

This method is applicable only to g modes that exhibit clear period-spacing variations, and ideally a larger number of modes.
We applied this technique to the g modes of SPB stars analyzed by \citet{Ped18} (Guo \& Aerts, in prep.), which exhibit a number of detected g modes ranging from $N = 6$ to $N = 23$. We found that the $FT(\Delta P)$ technique remains effective even for stars with a very limited number of g modes (as few as $N = 6$). However, misidentification or incompleteness of the $\Delta P$ series can significantly affect the results if the range of observed periods is not well covered. It should also be noted that with a limited number of modes, the classical approach of assigning error bars to frequencies in the Fourier spectrum based on the signal-to-noise ratio is no longer valid. Instead, a Monte Carlo or bootstrapping method should be employed to estimate the uncertainties.
 
Since we rely on the variational result to link pulsation perturbation
to the Brunt profile perturbation, the theoretical relations in Sec 2 and 5 apply only
 to g modes with small relative period perturbations $\delta P/P$.
This is usually the case for SPB and $\gamma$ Dor stars,
but may not hold for sdB and white dwarfs, which feature multiple chemical transition regions with sharper
 Brunt profile gradients.
For large-amplitude glitches, we can decompose the glitch into a series of small-amplitude
glitches, and apply the FT(DP) technique to each \citep{Zha23}.
Although this increases complexity - introducing harmonic peaks, etc.
We can mitigate these by using the pre-whitening or the CLEAN algorithm to remove spurious peaks.
This approach remains a promising tool for probing the chemical composition gradients in stellar interiors.

We only use the first-order asymptotic period relation in Eq. 3, which ignores the
second and high order terms. We have not fully exploited the information on the amplitudes of $FT(\Delta P)$,
i.e., Brunt profile derivatives from models.
Constructing denser grids of stellar models with varying mixing and overshooting parameters
can further constrain the desired stellar parameters.
Additionally, we did not consider rotating stellar models. A large number of 
$FT(\Delta P)$ amplitude measurements could potentially to constrain
rotational mixing in stellar structure computations,
and different convective-mixing schemes \citep{Nol24}.

Pulsational period spacings $\gamma$ Dor stars sometimes show
dips that are induced by the coupling between the pure inertial
mode in the convective and the gravity modes \citep{Saio21, Oua20}.
Our technique is not adequate for modeling these variations(dips)
in period spacings.

\textbf{(6)} On the core/envelope symmetry

\citet{Mon03} showed that buoyancy glitch $\delta N/N$ and its reflected version around
$u=0.5$ can induce the same g-mode period perturbations. They call it the inherent 
core/envelope symmetry which can lead to ambiguity in determining the location
 of features such as composition transition zones.
As we demonstrate earlier, this symmetry is a natural consequence of the normalized 
buoyancy radius $u$ being equivalent to the frequency $\xi$ and the 
symmetry around the Nyquist frequency $0.5$.

We summarize the procedures for analyzing observed g-mode periods $P_{k,obs}$ that exhibit small dips:
1. Identify the spherical degree $l$ and azimuthal order $m$.
 Fit the asymptotic relation to $P_k=\frac{\Pi_0}{\sqrt{\lambda_{l,m,s}}}(k + k_0)$  
to determine the asymptotic period spacing $\Pi_0$ and rotational frequency $f_{rot}$.
This step can be performed using the AMIGO package \citep{Van16} .
2. Convert the pulsation periods and period spacings to co-rotating frame,
and then compute the deviations of pulsation periods from the asymptotic
relation $\delta P$.
3. Perform a Fourier transform of the normalized $\Delta P_k$ and $\delta P/P$ as a function of 
radial order $k$ (the exact $k$ values are critical).
4. Identify the peak frequency $u_{\mu}$ 
and its amplitude in the Fourier spectra $FT(\Delta P_k)$. For stars near the terminal-age main sequence (TAMS), 
exercise caution when selecting the correct peak ($u_{\mu}$ can exceed $0.5$) and avoid 
spurious reflection peaks.

\begin{acknowledgements}
   We thank Conny Aerts, Tao Wu and Gang Li for helpful discussions. The research 
   leading to these results has received funding from the
   European Research Council (ERC) under the Horizon Europe
   programme (Synergy Grant agreement N$^\circ$101071505: 4D-STAR).
   While funded by the European Union, views and opinions expressed
   are however those of the author(s) only and do not necessarily
   reflect those of the European Union or the European Research
   Council. Neither the European Union nor the granting authority
   can be held responsible for them.
\end{acknowledgements}

%
\bibliographystyle{aa} 
\bibliography{dNdu} 
%



\end{document}